# SegQC: a segmentation network-based framework for multi-metric segmentation quality control and segmentation error detection in volumetric medical images

Bella Specktor-Fadida[1] PhD, Liat Ben-Sira[2,3] MD, Dafna Ben-Bashat[3,4,5] PhD, Leo Joskowicz[6,7] PhD

1. Department of Medical Imaging Sciences, University of Haifa, Israel

2. Division of Pediatric Radiology, Tel Aviv Sourasky Medical Center, Tel Aviv, Israel

3. Sackler Faculty of Medicine, Tel Aviv University, Israel

4. Sagol Brain Institute, Tel Aviv Sourasky Medical Center, Israel

5. Sagol School of Neuroscience, Tel Aviv University, Israel

6. School of Computer Science and Engineering, The Hebrew University of Jerusalem, Israel

7. Edmond and Lily Safra Center for Brain Sciences, The Hebrew University of Jerusalem, Israel

Email: bspecktor@univ.haifa.ac.il

**Abstract**

Quality control of structures segmentation in volumetric medical images is important for identifying segmentation errors in clinical practice and for facilitating model development by enhancing network performance in semi-supervised and active learning setups. This paper introduces SegQC, a novel framework for segmentation quality estimation and segmentation error detection. SegQC computes an estimate measure of the quality of a segmentation in volumetric scans and in their individual slices and identifies possible segmentation error regions within a slice. The key components include: 1) SegQC-Net, a deep network that inputs a scan and its segmentation mask and outputs segmentation error probabilities for each voxel in the scan; 2) three new segmentation quality metrics, two overlap metrics and a structure size metric, computed from the segmentation error probabilities; 3) a new method for detecting possible segmentation errors in scan slices computed from the segmentation error probabilities. We introduce a new evaluation scheme to measure segmentation error discrepancies based on an expert radiologist corrections of automatically produced segmentations that yields smaller observer variability and is closer to actual segmentation errors. We demonstrate SegQC on three fetal structures in 198 fetal MRI scans – fetal brain, fetal body and the placenta. To assess the benefits of SegQC, we compare it to the unsupervised Test Time Augmentation (TTA)-based quality estimation. Our studies indicate that SegQC outperforms TTA-based quality estimation in terms of Pearson correlation and MAE for fetal body and fetal brain structures segmentation. Our segmentation error detection method achieved recall and precision rates of 0.77 and 0.48 for fetal body, and 0.74 and 0.55 for fetal brain segmentation error detection respectively. SegQC enhances segmentation metrics estimation for whole scans and individual slices, as well as provides error regions detection.

## 1. Introduction

The segmentation of structures in volumetric medical images is increasingly used in clinical practice for a variety of diagnostic and prognostic tasks. Since manual delineation of structures' contours is time-consuming and requires expertise, a variety of automatic segmentation methods have been developed. In turn, the increased use of automatic segmentation methods creates the need for the systematic evaluation of the computed segmentation (Renard et al., 2020).

Automatic quality control (QC) of structures segmentations in volumetric scans can play an important role in clinical practice. It may provide an approach to systematically detect and correct segmentation errors that might go unnoticed, ensuring that the segmentations accurately represent the underlying anatomy. While segmentation quality assessment can be performed by a human expert, it is impractical for large datasets and is subject to observer variability and to interpretation and attention errors. Automatic quality estimation is key to validate the robustness of the segmentation methods on unseen scans.

Automatic segmentation QC can also be useful for the development of large datasets with high quality expert-validated annotations. Most state-of-the-art automatic segmentation methods for volumetric scans are based on deep neural networks that require a large, high-quality dataset of expert-validated annotations, which is expensive and very difficult to obtain (Tajbakhsh et al., 2020). A variety of methods have been recently developed to tackle segmentation annotation in medical images. Methods for leveraging unlabeled data to create better classification models with fewer annotated datasets, e.g. with self-training (Cheplygina et al., 2019) and active learning (Tajbakhsh et al., 2020) have been proposed. In the self-training setup, a model improves the quality of the pseudo-annotations by learning from its own high quality predictions. Active learning is an iterative paradigm wherein the unlabeled samples for each round of annotations are selected to maximally improve the performance of the current model. For both self-training and active learning, a QC mechanism is needed to select high-quality pseudo-labels and to identify low-quality cases for manual labeling.

Existing supervised methods for segmentation QC compute a single metric for the estimation, e.g., the Dice score. While the Dice score is usually the most technically relevant metric for the evaluation of segmentations, other metrics may be relevant for specific clinical tasks. For example, the Relative Volume Difference (RVD) is useful for quantifying volume measurement error. Moreover, the training of QC networks is usually performed using multiple annotated samples of varying quality, which requires large, annotated datasets and the training of multiple networks to produce appropriate segmentation masks. It is also important to evaluate QC methods on manual segmentations, which are usually of high quality for most of the structure except for a few isolated regions, which should be automatically identified and corrected. Furthermore, most methods focus on the precise estimation of the metric itself, while in practice the relative ranking of the segmentations by their quality or the identification of the cases with large segmentation errors is both preferable and sufficient.

Unsupervised methods for QC that do not require explicit supervision have also been proposed. They are used for ranking cases by quality for active cases selection, for pseudo-labeling, and for detecting outliers (Specktor-Fadida et al., 2023a; Audelan and Delingette, 2021). Unsupervised methods identify outliers in a set of segmentations, but may not be accurate in assessing individual segmentation results. Furthermore, they do not identify all erroneous segmentations in the dataset, but only some suspicious cases. Their advantage is that they do not require any supervision.

For segmentation QC, it is useful to identify segmentation error regions within scan slices to help annotators focus on relevant error regions that require manual correction. Several methods have been recently proposed to address the problem of error regions detection, e.g., Audelan and Delingette (2021); Zaman et al. (2023); Qiu et al. (2023). However, they do not provide both segmentation error regions detection and multiple quality metrics estimations in 2D and 3D in a single framework.

Ground truth quality is usually defined as the segmentation evaluation metric, e.g. the Dice score, between the ground truth segmentation and the computed segmentation. However, the discrepancy between the ground truth and predicted mask includes both the observer variability and the segmentation error (Joskowicz et al., 2019). This issue is usually overlooked when evaluating segmentation performance. Correcting segmentation masks introduces inductive bias to the manual segmentation and may result in reduced observer variability compared to manual delineations from scratch. Thus, evaluating QC performance on corrections data may be a more accurate error assessment.

This paper introduces SegQC, a novel framework for segmentation quality estimation based on overlap and size metrics and for segmentation error detection in volumetric scan slices. The key observation is that a deepl learning network, called SegQC-Net, can correctly identify the segmentation error regions in high quality segmentations despite the poor error segmentation performance. In addition, since the observer variability occurs in the boundaries of the structure of interest, it can be estimated using an offset band around the segmentation mask contour. We leverage these observations to design a quality estimator that can also identify error regions. We demonstrate the performance of our method on settings with different qualities and on corrections data and compare it to unsupervised quality estimation based on Test Time Augmentations (TTA) to see the benefits of the supervised approach (Dudovitch et al., 2020; Specktor-Fadida et al., 2021, 2023a).

## 2.  Related work

Multiple supervised methods have been proposed for segmentation QC that estimates a segmentation score. Reverse Classification Accuracy (RCA) has been proposed for QC assessment without the need of a large annotated dataset (Valindria et al., 2017; Robinson et al., 2019). However, both require long computation times, which preclude their use in real-time applications. Recent methods for quality control based on deep learning input a scan and a segmentation mask and output a predicted Dice coefficient (Robinson et al., 2018; Fournel et al., 2021), or the Intersection over Union (IoU, Jaccard index) with a regression network (Arbelle et al., 2019; Huang et al., 2016; Shi et al., 2017). Liu et al. (2019) use a Variational Auto-Encoder (VAE) network to learn shape features and a regressor in the one-dimensional feature space to predict the quality of the segmentation. These methods output quality scores but are limited to the prediction of a single predefined quality metric and require large datasets of scan-mask pairs with masks of varying quality. Furthermore, they usually provide a single score for the entire scan, not for individual slices where the manual corrections should be made.

Unsupervised methods have been proposed for segmentation QC to alleviate the burden of data annotation. Methods for unsupervised quality control rely on assumptions about the appearance and shape of the structure of interest and of the background regions, e.g. high levels of intra-region homogeneity and inter-region heterogeneity (Rosenberger et al., 2006; Chabrier et al., 2006; Zhang et al., 2008; Johnson and Xie, 2011; Gao et al., 2017). Audelan and Delingette (2021) proposed unsupervised quality control where quality estimates are produced by comparing each segmentation with the output of a probabilistic segmentation model that relies on intensity and smoothness assumptions. Ranking cases with respect to these two assumptions allows the detection of the most challenging cases in a dataset. Other works use TTA based Dice score estimation based on the Dice coefficient between each one of the augmentation results and the average or median prediction Dudovitch et al. (2020); Specktor-Fadida et al. (2021, 2023a,b). These methods were used for scans prioritization in active learning, high quality pseudo-labels selection in self-training, and for discarding low quality segmentations when estimating weight from fetal MRI whole-body segmentation masks.

Other methods use the uncertainty of segmentations to assess their quality. Uncertainty quantification also adds interpretability to the quality assessment as it provides information about the location of possible errors. Roy et al. (2019) uses Monte Carlo dropout to estimate uncertainty, which yields a good correlation between the measured uncertainty and the Dice coefficient. In DeVries and Taylor (2018), a first network computes a segmentation map and an uncertainty map at the pixel level, which are then used by a second network to regress a QC estimate at the image level. Dudovitch et al. (2019) describe a TTA-based entropy uncertainty measure that is used to prioritize slices for correction.

Several methods have been proposed for the detection of segmentation error regions. Zaman et al. (2023) predict segmentation error regions with a 3D segmentation network where the ground truth error is the absolute difference between the ground truth and the predicted masks. They then extract erroneous surface patches with the Marching Cubes and k-d tree algorithms. Qiu et al. (2023) use a combined segmentation and regression network to simultaneously compute segmentation quality measures and voxel-level segmentation error maps for brain tumor segmentation quality control. However, this method computes a single Dice score measure in 3D for each scan and does not provide 2D information. It uses the Dice score to quantify error regions detection, which may not be optimal for cases with correct detected error regions but failure of precise segmentation.

The discrepancy between ground truth segmentation and predicted segmentation mask includes the observer variability and the segmentation error (Joskowicz et al., 2019). Correcting the computed segmentation mask introduces inductive bias to the manual segmentation and may result in reduced observer variability compared to manual delineations from scratch (Chlebus et al., 2019). However, to the best of our knowledge, none of the existing methods have been tested on segmentation masks correction data. In our previous work (Specktor-Fadida et al., 2022), we used corrections data to prioritize slices for manual correction and tested the method on fetal body corrections test data. However, this method required a dedicated dataset of correction data for training, and it did not perform metrics estimation, 3D scans prioritization and error detection.

## 3. Method

We describe next SegQC, a novel method for simultaneously computing quality estimates for the entire scan and individual slices using overlap and size metrics and identifying segmentation error regions. It estimates the Dice score and the IoU overlap metrics, and the absolute Relative Volume Difference (ARVD) for volume estimation. It requires a relatively small number of annotated cases for training and uses input-mask pairs generated from a single segmentation network.

The SegQC segmentation quality estimation framework uses SegQC-Net, a network to predict segmentation error regions, followed by quality estimation modules using the segmentation results (Fig. 1). The segmentation error computed by the SegQC-Net on a given scan-mask is used to compute estimated 2D and 3D Dice scores and ARVD metrics. To detect and highlight segmentation error regions, an estimated error is computed from the network's output mask and error regions are calculated from the estimated error result. Finally, scans and slices are ranked using the SegQC-Net output for scans prioritization based on largest error.

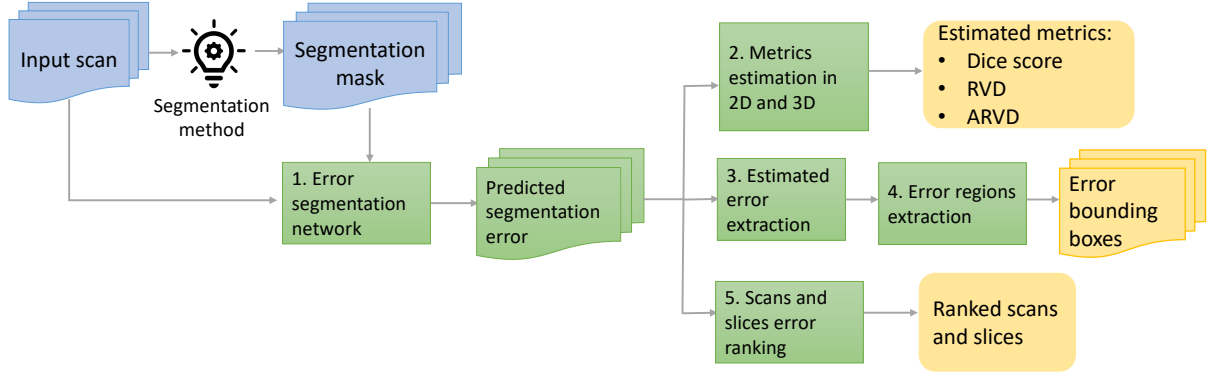

**Fig. 1.** The SegQC segmentation quality estimation method. The segmentation error estimation proceeds as follows: 1) segmentation error computation with SegQC-Net, a U-Net that inputs the scan and a segmentation mask from the previous step; 2) Dice score and Absolute Relative Volume Difference (ARVD) estimations in 2D and 3D from the predicted segmentation error result; 3) estimated segmentation error extraction from the predicted segmentation mask; 4) segmentation error regions extraction from the estimated error result. Inputs are colored in blue, outputs in yellow; 5) 3D scans and 2D slices ranking from largest to smallest predicted segmentation errors.

### 3.1. Segmentation error estimation network

The segmentation error estimation network, SegQC-Net, inputs a volumetric scan $I \in R^{n \times m \times k}$ and the segmentation mask $M \in N^{n \times m \times k}$. It outputs the segmentation error in the mask $E \in R^{n \times m \times k}$, where $m, n, k \in N$ are the dimensions of the volumes.

The training data for the SegQC-Net is created as follows (Figure A1, Supplemental Material). First, to produce scan-mask pairs, a segmentation network is trained to create the segmentation masks. To increase the number of scan-mask pairs, Test-time Augmentations (TTA) are used.

Segmentation masks augmentations are performed to increase the number of cases with ground truth segmentation delineations and to have pairs of input segmentation masks with different qualities to account for the high variability of possible segmentation errors. We create multiple masks with the same input using TTA on the segmentation network (Figure A1, Supplemental Material). These constitute mask augmentations. The final input-mask data consists of multiple masks for each input, as well as the median TTA prediction. For TTA, we perform flipping, rotation, transpose and contrast augmentations.

### 3.2. Metrics estimation

The Dice score, IoU and ARVD metrics are computed as follows. Let $t \in T$ be a voxel in the ground truth segmentation $T$, $m \in M$ be a voxel in the segmentation mask $M$ and $i$ be an index over the voxels in the volume. Then:

$$(1) \quad Dice = \frac{2 \sum_i^M m_i t_i}{\sum_i^M m_i + \sum_i^M t_i}$$

$$(2) \quad IoU = \frac{\sum_i^M m_i t_i}{\sum_i^M m_i + \sum_i^M t_i - \sum_i^M m_i t_i}$$

$$(3) \quad ARVD = \frac{\sum_i^M |m_i - t_i|}{\sum_i^M t_i}$$

Note that the ground truth segmentation $T$ is unknown. Given a segmentation mask and the result of an error segmentation network, we estimate the Dice score and RVD quality metrics of the masks as follows. Let $M$ be the segmentation mask, $\hat{E} \in (0,1) \subset R$ be the estimated error from the error segmentation network result and $\epsilon$ be the smoothing term. We first compute the estimated ground truth $\hat{T} \in (0,1) \subset R$ by summing the estimated intersection with mask $M$ and the estimated truth which is a segmentation error in mask $M$:

$$(4) \quad \hat{T} = M \cdot \left(1 - \hat{E}\right) + (1 - M) \cdot \hat{E}$$

Let $\hat{t} \in \hat{T}$ be a voxel in volume $\hat{T}$, $m \in M$ a voxel in mask $M$ and $\hat{e} \in \hat{E}$ a voxel in volume $\hat{E}$. Since $M \cdot (1 - \hat{E})$ is the intersection between mask $M$ and ground truth, the estimated Dice is:

$$(5) \quad Dice_{est} = \frac{2 \sum_i^M m_i \cdot (1 - \hat{e}_i)}{2 \sum_i^M m_i + \sum_i^M \hat{t}_i}$$

The estimated $RVD$ and $ARVD$ are:

$$(6) \quad RVD_{est} = \frac{\sum_i^M (m_i - \hat{t}_i)}{\sum_i^M \hat{t}_i}$$

$$(7) \quad ARVD_{est} = \frac{\sum_i^M |m_i - \hat{t}_i|}{\sum_i^M \hat{t}_i}$$

Note that the denominator is the segmentation error difference, so it can be estimated directly by the network output. Thus, we can simplify the equation:

$$(8) \quad ARDV_{est} = \frac{\sum_i^M \hat{e}_i}{\sum_i^M \hat{t}_i}$$

### 3.3. Estimated error extraction

Ideally, the ground truth segmentation error should only correspond to regions with segmentation errors and should exclude regions of observer variability. Formally, let $D_t \in N^{n \times m \times k}$ be the difference volume between a given segmentation mask $M \in N^{n \times m \times k}$ and the ground truth segmentation $T \in N^{n \times m \times k}$, and $V_t \in N^{n \times m \times k}$ be the observer variability regions. The ground truth segmentation volume error $E_t \in R^{n \times m \times k}$ is:

$$(9) \quad E_t = T - M - V_t = D_t - V_t$$

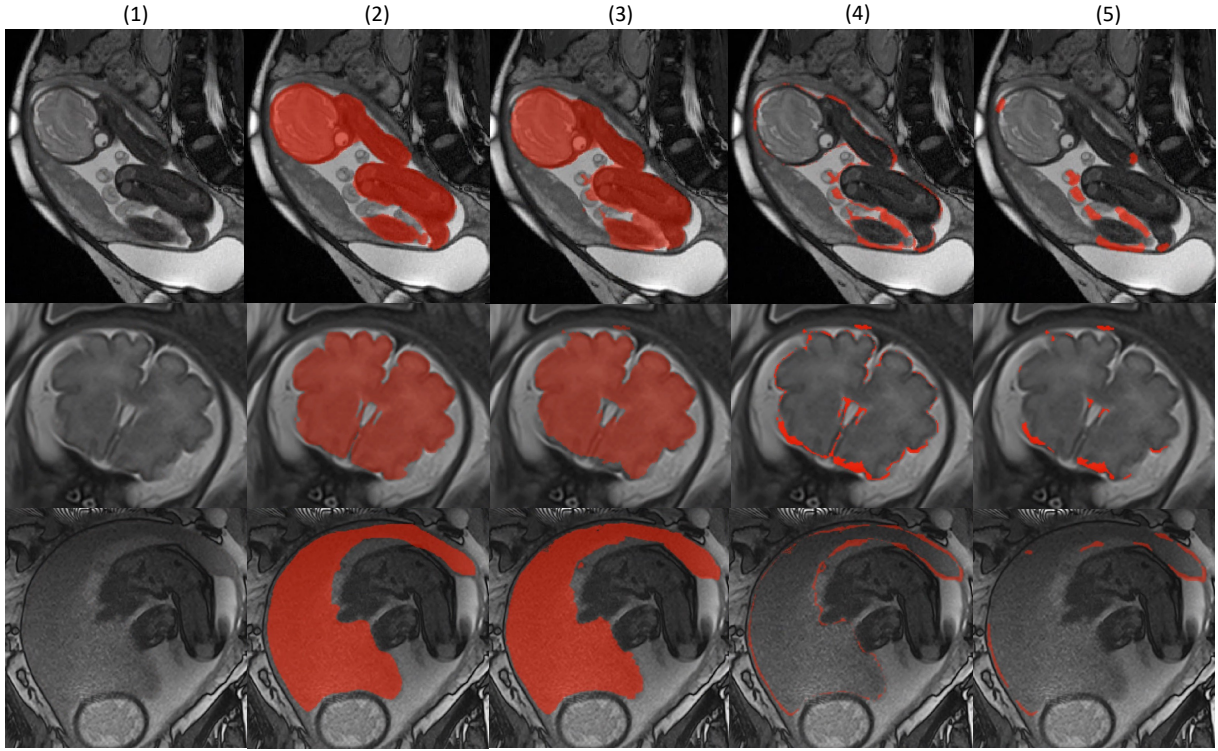

**Fig. 2.** Illustration of the estimated error extraction algorithm for the fetal body, brain and placenta segmentation tasks. 1) input image; 2) ground truth segmentation; 3) segmentation mask; 4) difference volume between ground truth and segmentation mask; 5) estimated error of the segmentation mask.

In practice, only $D_t$ can be computed and not $E_t$, as the observer variability mask $V_t$ is not available. However, when the ground truth segmentation was created by manually correcting the segmentation mask $M$, the mask $M$ reflects the inductive bias of the manual delineation and can help reduce the observer variability $V_t$, thereby reflecting the segmentation error $E_t$.

Ideally, the goal is to obtain multiple scan mask pairs with ground truth segmentations that were created by correcting the input mask. However, it is very difficult to obtain many cases with segmentation corrections data. Furthermore, we would like to train our error segmentation network with masks of both high and low quality, although corrections datasets are obtained with high quality segmentations, for otherwise the manual delineations would have been performed from scratch. Therefore, instead of using real segmentation correction data, we estimate the segmentation error regions from the ground truth difference volume $D_t$. The segmentation error estimation algorithm uses the simple heuristic of removing a band around the difference volume $D_t$ contour. The idea behind it is that much of the variability $V_t$ lies in the boundary of the structure of interest, thus to estimate the segmentation error boundary can be removed.

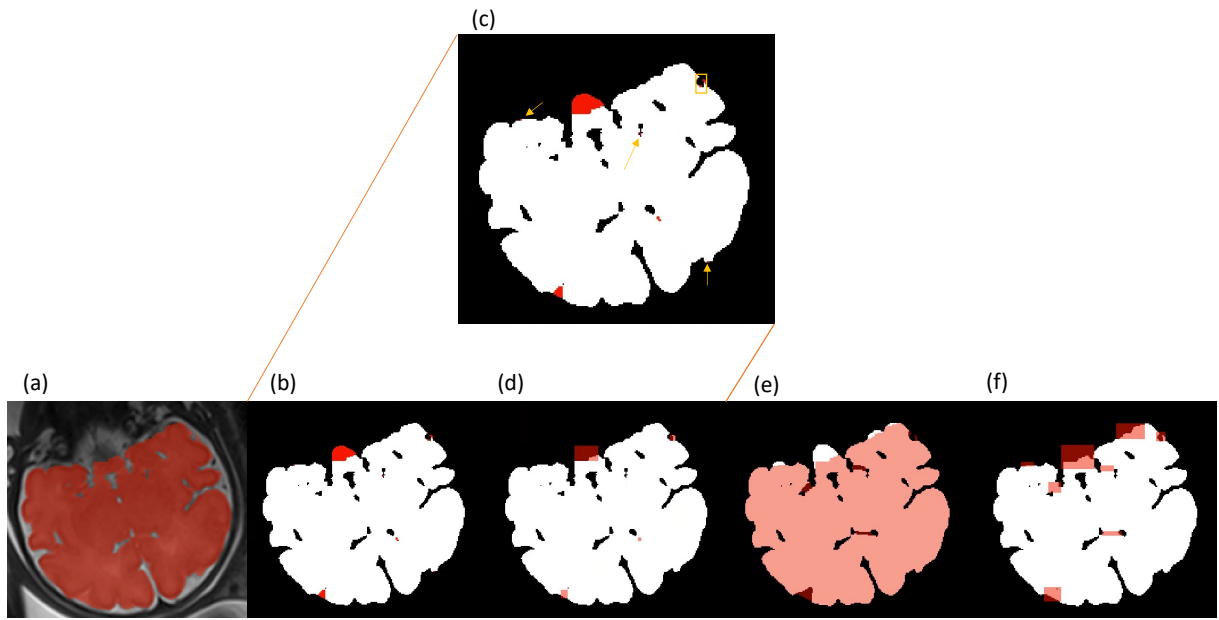

**Fig. 3.** Illustration of the boxes ROI extraction. (a) Input mask for error evaluation; (b) Estimated error output; (c) Estimated error postprocessing: removed small error regions are marked in yellow arrows; unified boxes are marked in yellow box; (d) Estimated error bounding boxes; (e) Overlay of the mask with ground truth mask; (f) Ground truth error bounding boxes.

The algorithm consists of three steps performed on each scan slice separately: 1) binary dilation and binary erosion; 2) removal of the area between dilation and erosion; 3) removal of small components below a threshold $t_{size}$. Fig. 2 shows examples for the fetal body, the fetal brain and the placenta.

*3.4. Segmentation error regions computation*

To extract segmentation error regions as 2D regions of interest (ROI) bounding boxes from the computed segmentation error, the following operations are applied: 1) extraction of the connected components; 2) extraction of bounding boxes around the connected components; 3) unification of the nearby boxes; 4) removal of boxes with small areas below a threshold. Fig. 3 illustrates this process.

*3.5. Scans and slices segmentation quality ranking*

Prioritizing segmentation corrections can be useful for different tasks, such as active learning. Segmentation volumes are prioritized for correction based on the chosen estimated metrics in 3D. Since 2D slices might not include the structure of interest, it is not clear how to prioritize between those slices using 2D metrics estimation. Thus, 2D corrections prioritization is performed using the summation function over the error segmentation network output in the slices.

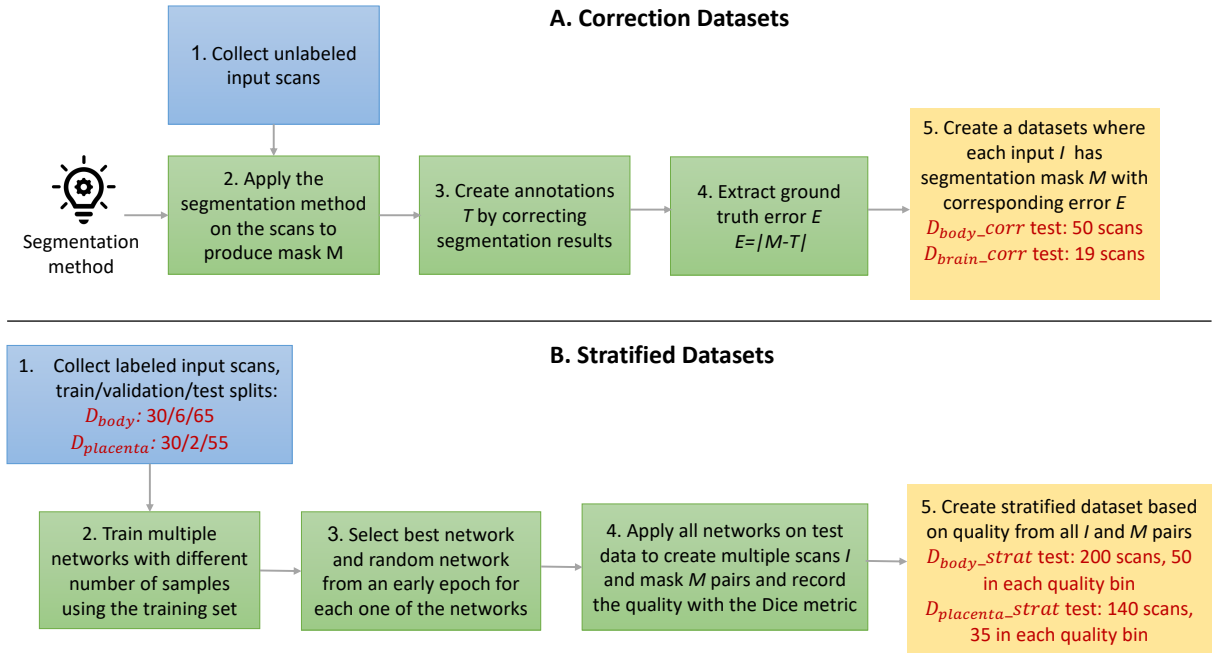

**Fig. 4**: The creation process of the test sets for corrections and stratified datasets. The number of the final test set cases is shown in red in the yellow output boxes.

## 4.  Experimental results

The experimental results are presented as follows. Section 4.1 provides a detailed description of the datasets used and the process of their construction. Section 4.2 outlines the specifics of the trained model. Section 4.3 explains the evaluation setup. Section 4.4 presents the experimental studies, and the results on various datasets and of the ablation studies.

### 4.1. Datasets

To evaluate our method, we used two types of data. The first is the manual segmentation correction data, **D_corr**, which includes the scan and segmentation mask pairs with their corresponding corrected segmentation by a human annotator. This data type is particularly useful to quantify error prediction, as data corrected from a given segmentation mask is subject to lower observer variability (Chlebus et al., 2019). The drawback of the segmentation corrections data is that it contains data with similar and usually high qualities, since in cases of a poor quality it is better to perform the manual segmentation from scratch. The second data type is segmentation data with stratified qualities, **D_strat,** as was suggested in previous works Robinson et al. (2018); Fournel et al. (2021). The process for creating the **D_corr** and **D_strat** test sets is depicted in Figure 4, while the creation of the training sets is presented in Figure A1 of the Supplemental Material.

**Corrections datasets**. Correction datasets were created from data for which the ground truth $T$ was created by correcting a segmentation mask $M$. The ground truth error $E$ was then computed using the absolute difference between a corrected segmentation $T$ and a segmentation mask $M$ from which the manual correction was performed, $E = |T - M|$.

**Stratified datasets**. To produce masks with variable qualities, the segmentation data was first split into training/validation/test. Next, 12 networks were trained with different subsets of the training data, with and without anisotropic pooling. To enlarge the number of low-quality masks, two network weights were selected from each network training session: networks weight that achieved the best results on the validation set and the network weights of a random early epoch. This resulted in a total of 24 networks weights. The networks were then used to compute predictions on the test set and the Dice score for each case was recorded. Then, for each Dice score bin (e.g. Dice score between 0.85 and 0.9) scan-mask pairs were selected with a higher priority to new scans that were not already in the bin. The same number of cases were selected for each quality bin. The final dataset was constructed with the selected stratified scan-mask pairs and their corresponding ground truth discrepancy $E$ between the segmentation mask $M$ and the manual annotation $T$, $E = |T - M|$.

We evaluated the segmentation quality estimation method on three fetal MRI structures: 101 fetal body, 40 fetal brain and 57 placenta cases. All scans were retrospectively obtained from Tel Aviv Sourasky Medical Center (Tel Aviv, Israel) by the co-author radiologist during routine clinical practice. For each one of the segmentation structures, we used corrections data when applicable and constructed stratified qualities datasets for structures with at least 50 annotated cases. Figure 4 summarizes the test sets details of all datasets.

Dataset $\boldsymbol{D_{body}}$ consists of 101 cases with the TRUFI sequence acquired on Siemens Skyra 3T, Prisma 3T, and Aera 1.5T scanners. Each scan has a resolution in the range of $0.6 - 1.34 \times 0.6 - 1.34 \times 2 - 4.8\ mm^3$. Two sub-cohorts were created, one with corrections data $\boldsymbol{D_{body\_corr}}$ and another using segmentation masks having stratified qualities $\boldsymbol{D_{body\_strat}}$:

- $\boldsymbol{D_{body\_corr}}$ consists of 101 scan-mask pairs $I$, $M$. Data was split to 25/26/50 training/validation/test. The test set had ground truth error $E$ that is the absolute difference between a corrected segmentation $T$ and a segmentation mask $M$ from which the manual correction was performed.

- $\boldsymbol{D_{body\_strat}}$ consists of scan-mask pairs $I$, $M$ with test-set masks that were created with stratified qualities. Data was split to 30/6/65 training/validation/test. Test set data was stratified with Dice score bins of $[0,0.88]$, $[0.88,0.92]$, $[0.92,0.96]$, $[0.96,1]$. Note that the first bin includes variable qualities, since the fetal body is a large structure usually with high quality segmentations and there were not enough samples for lower quality bin threshold. The stratified test set resulted in 200 cases, 50 cases in each bin.

Dataset $\boldsymbol{D_{brain}}$ was used for fetal brain segmentation quality estimation. It consists of 40 cases of the HASTE sequence acquired on Siemens Skyra 3T, Prisma 3T, and Aera 1.5T scanners with resolution in the range of $0.45 - 1.5 \times 0.45 - 1.5 \times 2.2 - 6 \ mm^3$. Out of the 40 cases, 19 cases were created by correcting segmentation results. Since the brain segmentation dataset was relatively small, we created only corrections dataset $\boldsymbol{D_{brain\_corr}}$ that included 19 cases for training, two for validation and 19 cases with corrections for testing.

Dataset $\boldsymbol{D_{placenta}}$ was used for placenta segmentation quality estimation. It consists of 57 placenta segmentation cases with the TRUFI sequence acquired on Siemens Prisma 3T and Vida 3T scanners with a resolution of $0.781 \times 0.781 \times 2 \ mm^3$. $\boldsymbol{D_{placenta\_strat}}$ was created from the data, a set of scan-mask pairs $I$, $M$ with test set masks that were created with stratified qualities. Data was split to 30/2/25 training/validation/test. Test set data was stratified with the Dice score bins of [0,0.75], [0.74,0.833], [0.833,0.917], [0.917,1]. The stratified test set resulted in 140 cases, 35 cases in each bin.

### 4.2. Model

The volumetric segmentation model SegQC-Net inputs a scan and a segmentation mask and outputs an error segmentation mask. It has an anisotropic network architecture similar to that of Dudovitch et al (2020). The model was trained with a Dice loss function, initial learning rate of 0.0005 and a batch size of 2. To normalize for the scan dimensions, block sizes of 256×256×48, 128×128×32 and 128×128×48 were used for fetal body, fetal brain, and placenta error segmentation networks respectively.

### 4.3. Evaluation

We used the Dice, IoU and AVDR metrics for the evaluation and the Mean Average Error (MAE) and Pearson correlation, as in Fournel et al. (2021); Qiu et al. (2023). We determined whether the MAE differences between methods are significant using paired two-sided t-tests.

We performed corrections evaluation in 3D and 2D to quantify the ability of correcting segmentations of volumes or 2D slices in descending order of quality, from the lowest to the highest quality. For 3D segmentation corrections evaluation, we evaluated the test scans correction order and compared it to the order based on TTA quality estimation, to the optimal order and to random order. For 2D segmentation corrections evaluation, we evaluated different correction percentages and compared them to 2D TTA quality estimation based on entropy, optimal order, random order, random order with non-empty slices and sequential order.

For 2D segmentation error detection evaluation, we first extracted bounding boxes around ground truth error and binarized prediction by first extracting bounding boxes around 2D connected components and then unifying boxes whose distance between them was smaller than a predefined minimum distance $min_d$. Then, we used the Average Precision (AP) and the Average Recall (AR) metrics to evaluate the detected bounding boxes compared to ground truth error bounding boxes with varying intersection-over-union percent- ages. We used small IoU percentages of 0.05-0.2 to account for the precise detection difficulty of small segmentation errors.

### 4.4. Experimental Studies

The Experimental Studies section is structured as follows. Section 4.4.1 presents the results on the corrections data. Section 4.4.2 discusses the outcomes in stratified datasets. Section 4.4.3 details the ablation studies.

### 4.4.1. Studies on corrections datasets

We evaluated our method on fetal body and fetal brain segmentation corrections datasets with metrics estimation, corrections in 3D and 2D based on quality ranking, and 2D detection evaluations. The fetal body test set consisted of 45 cases; the fetal brain consisted of 19 cases. All cases were created by correcting the results of same network for each structure. The training of error segmentation networks was performed on segmentation masks created from these two networks. The results were compared to TTA-based quality estimation using the same network from which the results were corrected from.

Table 1 shows MAE of metrics estimation results. Tables A1 and A2 (Supplemental Materials) show statistical p-values for fetal body and brain segmentation corrections datasets. Table 2 lists the Pearson correlation results. For both fetal body and fetal brain segmentation quality estimation and for all metrics, the Pearson correlations are higher for the SegQC network compared to TTA estimation, with and without error extraction.

For fetal body segmentation quality estimation, the MAE results of 2D metrics are significantly better using the SegQC network compared to TTA-based estimation. For fetal brain segmentation, for all metrics except for 3D AVDR, the MAE results are significantly better for the SegQC estimation compared to TTA-based quality estimation method.

| | | Corrections data | | | | Stratified data | | | |
|---|---|---|---|---|---|---|---|---|---|
| | | Body | | Brain | | Body | | Placenta | |
| | | TTA | SegQC | TTA | SegQC | TTA | SegQC-TR | TTA | SegQC-TR |
| **3D** | Dice | 0.013 | **0.008** | 0.052 | **0.020*** | 0.056 | **0.041*** | 0.160 | **0.082*** |
| | IoU | 0.021 | **0.015** | 0.084 | **0.033*** | 0.081 | **0.064*** | 0.228 | **0.104*** |
| | AVDR | 0.020 | **0.018** | **0.057** | 0.061 | 0.062 | **0.057** | **0.185*** | 0.251 |
| **2D** | Dice | 0.028 | **0.023*** | 0.084 | **0.050*** | 0.073 | **0.058*** | 0.137 | **0.111*** |
| | IoU | 0.044 | **0.038*** | 0.117 | **0.066*** | 0.097 | **0.078*** | 0.190 | **0.129*** |
| | AVDR | 0.043 | **0.040*** | 0.104 | **0.079*** | 0.103 | **0.095*** | **0.205*** | 0.274 |

**Table 1.** Segmentation quality estimation Mean Average Error (MAE) for fetal body and brain corrections data and fetal body and placenta stratified datasets. TTA - test time augmentations, SegQC - SegQC difference network trained on masks pairs from corrected segmentation network, SegQC-TR - SegQC network using difference network trained on masks of a single network. Significant values with p<0.01 are marked with *.

| | | Body | | | Brain | | |
|---|---|---|---|---|---|---|---|
| | | TTA | SegQC | SegQC-EE | TTA | SegQC | SegQC-EE |
| **3D** | Dice | 0.229 | 0.692 | **0.699** | 0.230 | 0.898 | **0.901** |
| | IoU | 0.349 | 0.697 | **0.703** | 0.322 | 0.914 | **0.915** |
| | AVDR | 0.243 | 0.521 | **0.529** | 0.420 | 0.601 | **0.605** |
| **2D** | Dice | 0.495 | 0.621 | **0.630** | 0.529 | **0.768** | 0.766 |
| | IoU | 0.530 | 0.608 | **0.619** | 0.557 | **0.812** | 0.811 |
| | AVDR | 0.497 | 0.618 | **0.631** | 0.483 | **0.718** | 0.718 |

**Table 2.** Fetal body and fetal brain segmentation quality estimation Pearson correlation for corrections data. TTA - test time augmentations, SegQC - SegQC method using difference network, SegQC-EE - SegQC method using difference network following error extraction.

Figures 5 and 6 show 3D and 2D correction results for the body and brain structures respectively. Despite the large differences in MAE and Person correlation for the fetal body segmentation error estimation, the differences in 3D and 2D segmentation correction results are minor. For the fetal brain segmentation error estimation, the differences in correction performance are larger, especially for 2D, but still not very large.

Segmentation error detection in 2D slices was also evaluated for segmentation errors with an area > 100 mm$^2$. For both fetal body and fetal brain segmentation error detection evaluation, we used $min_d$ = 5 voxels to unify bounding boxes and a threshold of $th$ = 0.5 for the network results. Very large bounding boxes, i.e., bounding boxes whose area was larger than half the area of the whole slice and larger than twice the area of the largest ground truth bounding box were discarded, as they are not useful for focusing on the specific error region.

Table 3 shows 2D detection results for fetal body and fetal brain segmentation errors > 100 mm$^2$. For fetal body segmentation error detection, using SegQC-Net followed estimated error extraction performed better compared to SegQC-Net without it, with a precision of 0.48 and a recall of 0.77 for IoU of 0.05% vs. a precision of 0.29 and a recall of 0.55. Computing the estimated segmentation errors from the SegQC-Net output was also beneficial for fetal brain segmentation error detection, with a precision of 0.55 and a recall of 0.74 vs. a precision of 0.52 and a recall of 0.63 without estimated error extraction.

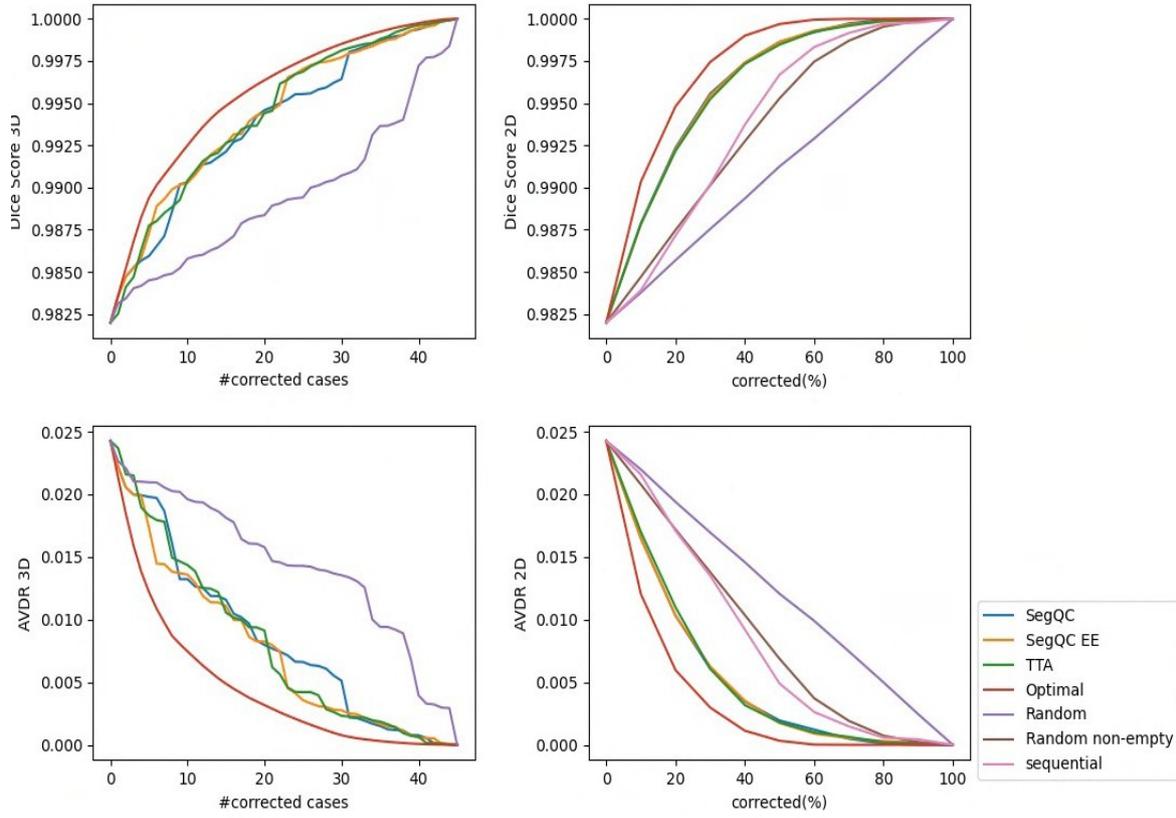

**Fig. 5.** Correction results based on masks ranking for fetal body segmentation quality estimation tasks.

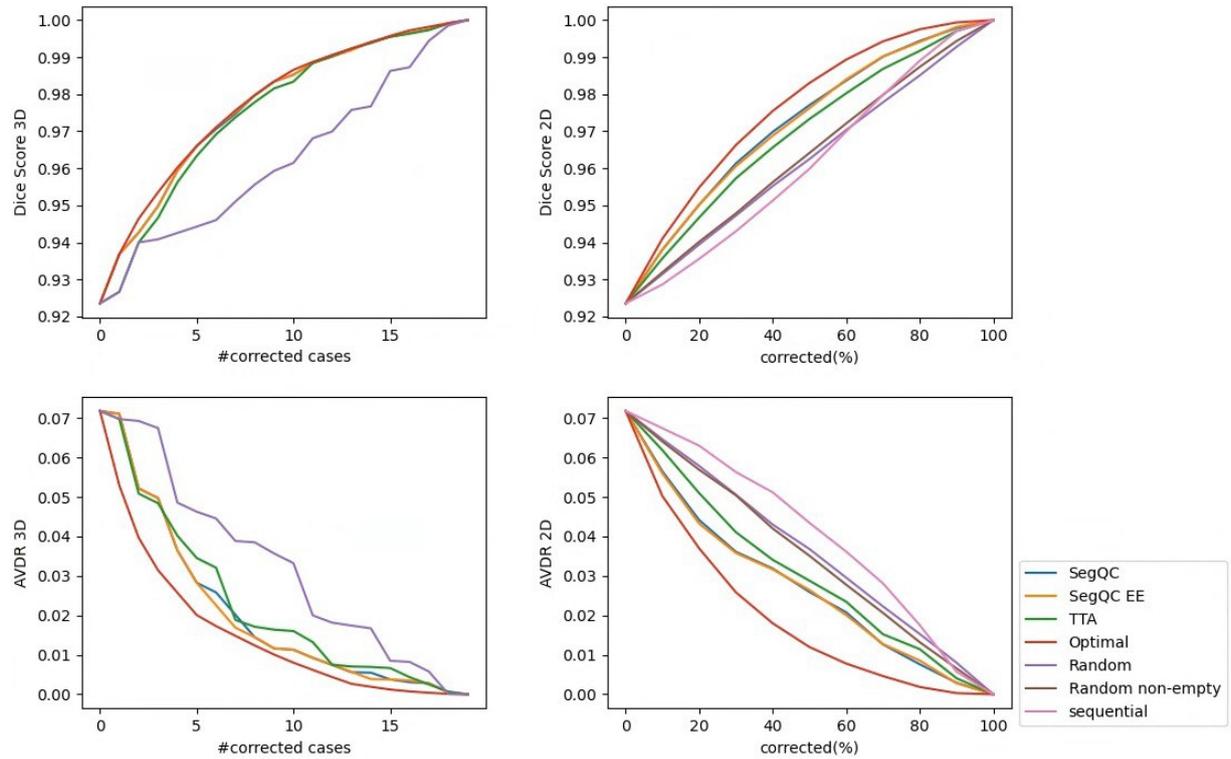

**Fig. 6.** Correction results based on masks ranking for fetal brain segmentation quality estimation tasks.

|  | IoU (%) | Body | | Brain | |
|---|---|---|---|---|---|
|  |  | \wo EE | \w EE | \wo EE | \w EE |
| **Precision** | 0.05 | 0.29 | **0.48** | 0.52 | **0.55** |
|  | 0.10 | 0.17 | **0.43** | 0.49 | **0.53** |
|  | 0.15 | 0.11 | **0.37** | 0.48 | **0.50** |
|  | 0.20 | 0.08 | **0.32** | 0.45 | **0.46** |
| **Recall** | 0.05 | 0.55 | **0.77** | 0.63 | **0.74** |
|  | 0.10 | 0.40 | **0.68** | 0.61 | **0.69** |
|  | 0.15 | 0.31 | **0.60** | 0.59 | **0.65** |
|  | 0.20 | 0.25 | **0.53** | 0.56 | **0.62** |

**Table 3.** Detection results for fetal body and fetal brain segmentation errors with average precision (AP) and average recall (AR) evaluation metrics for different IoU %. Comparison between SegQC-Net with (\w EE) and without (\wo EE) error extraction.

### 4.4.2. Studies on stratified datasets

We evaluated our method on stratified fetal body and placenta datasets with metrics estimation and segmentation corrections in 3D and 2D based on quality ranking. Detection evaluation was not feasible for stratified datasets as we did not have ground truth corrections data for these cases. Results of the SegQC method were compared to TTA-based quality estimation using the same networks from which the evaluated segmentation masks were created from. To evaluate the transferability of the SegQC-Net segmentations, we compared training of a SegQC network on masks using stratified masks qualities to using masks of only one of the segmentation networks that produces relatively high-quality masks.

Tables 5 and 7 list the metrics estimation MAE and Pearson correlation respectively, and Fig. 7 shows corrections results for the fetal body stratified dataset. Metrics estimation in terms of MAE are significantly better using SegQC network compared to TTA estimation except for the 3D AVDR metric when training with stratified data. Pearson correlations for the SegQC network are higher compared to TTA-based estimation. Furthermore, metrics estimation using the SegQC network trained on masks of a single network without stratification outperformed TTA metrics estimation as well. However, in terms of correction graphs in both 3D and 2D there is almost no difference between ranking using TTA estimation and ranking using SegQC networks masks qualities to using masks of only one of the segmentation networks that produces relatively high-quality masks.

| | | Body | | | Placenta | | |
|---|---|---|---|---|---|---|---|
| | | TTA | SegQC-TR | SegQC | TTA | SegQC-TR | SegQC |
| **3D** | **Dice** | 0.523 | 0.667 | **0.807** | 0.633 | 0.869 | **0.901** |
| | **IoU** | 0.641 | 0.791 | **0.885** | 0.682 | 0.850 | **0.881** |
| | **AVDR** | 0.469 | 0.573 | **0.626** | 0.407 | 0.566 | **0.624** |
| **2D** | **Dice** | 0.614 | 0.707 | **0.831** | **0.701** | 0.680 | 0.655 |
| | **IoU** | 0.708 | 0.768 | **0.874** | **0.736** | 0.690 | 0.675 |
| | **AVDR** | 0.640 | 0.654 | **0.795** | 0.340 | 0.461 | **0.466** |

**Table 4.** Pearson correlation results for fetal body and placenta segmentation quality estimation for stratified dataset (200 test examples). TTA - test time augmentations, SegQC-TR - SegQC method using difference network trained on masks of a single network for transferability test, SegQC - SegQC method using difference network trained on stratified masks.

Tables 1 and 4 list the metrics estimation MAE and Pearson correlation respectively. Fig. 8 shows the corrections results for placenta segmentation quality estimation on the stratified dataset. While 3D overlap metrics estimations are significantly better using the SegQC networks in terms of MAE and show improved Person correlations, 3D AVDR metric and 2D metrics demonstrate mixed results. Correction graphs showed slight benefit for ranking based on SegQC network compared to ranking based on TTA for 3D corrections, and similar performance for 2D corrections. Using SegQC networks, metrics estimation in 2D demonstrated a decline in performance compared to 3D metrics estimation.

Using SegQC-Net networks trained on masks using a single network compared to stratified masks usually resulted in a decline in metrics estimation performance, with larger difference in the case of fetal body segmentation quality estimation compared to placenta segmentation quality estimation. However, the difference in correction graphs performance was much smaller, with the largest difference observed for 3D fetal body corrections.

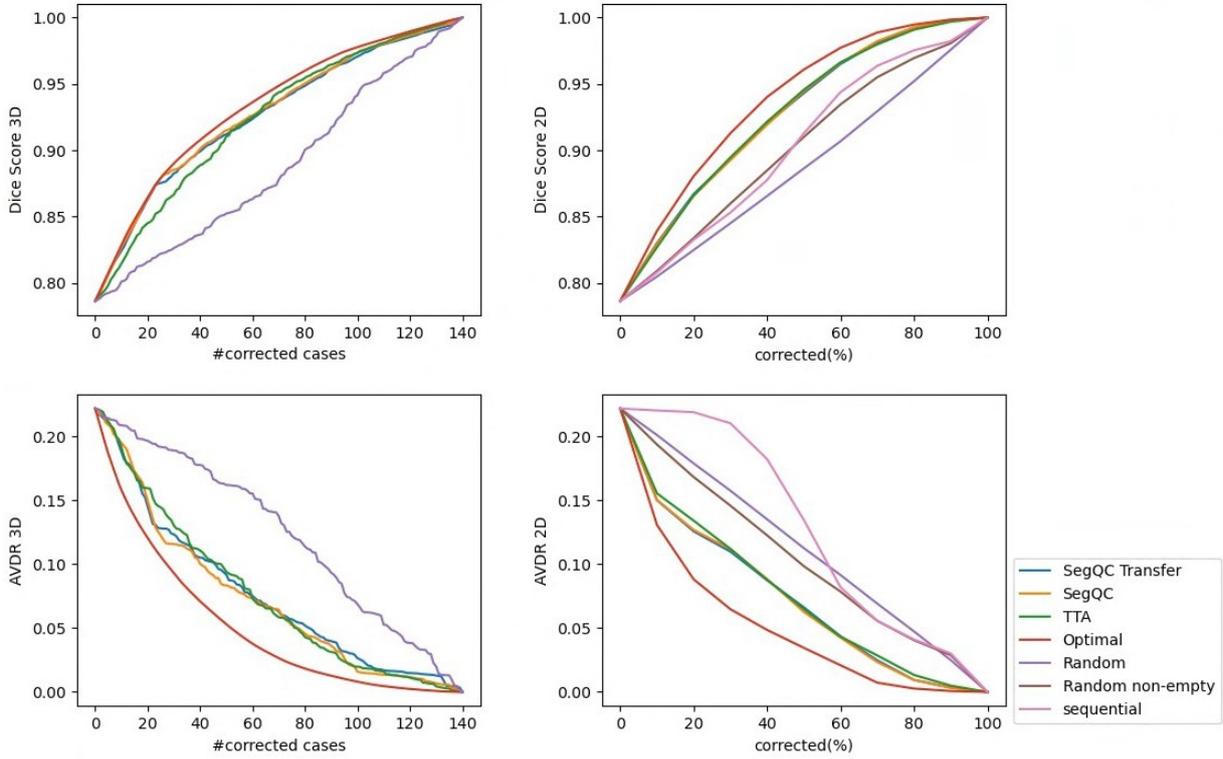

**Fig. 7.** Correction results for fetal body segmentation quality estimation using stratified dataset.

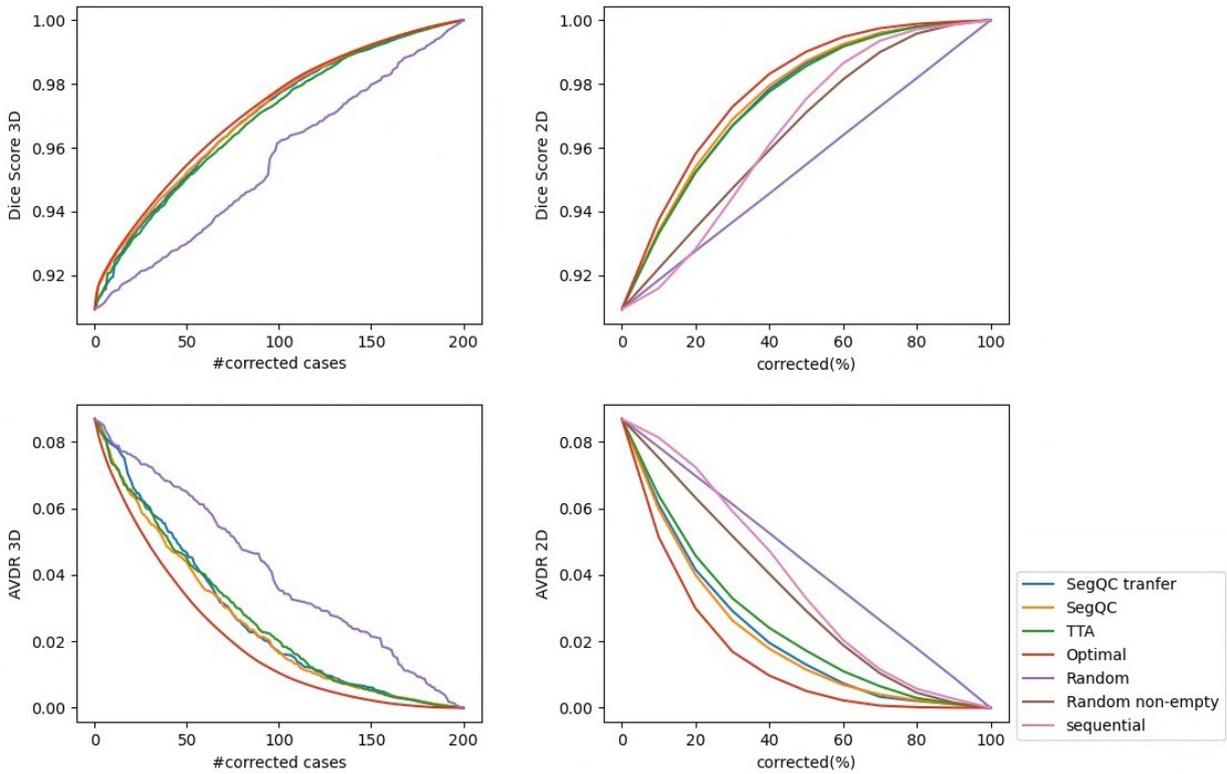

**Fig. 8.** Correction results for placenta segmentation quality estimation using stratified dataset.

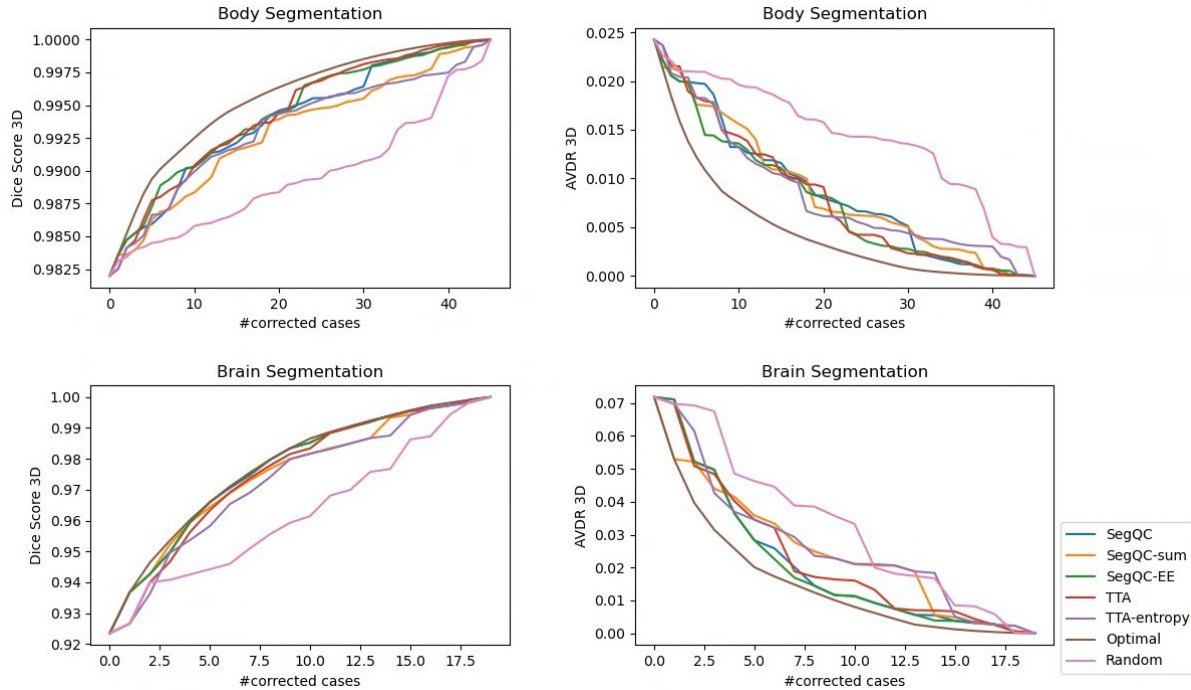

**Fig. 9.** Correction results for 3D data - comparison between metrics estimation using SegQC and TTA and sum and entropy functions respectively.

### 4.4.3. Ablation studies

**Metrics estimation ablation for 3D corrections evaluation**. Cases prioritization for correction was performed based on 3D metrics estimation ranking. To understand the benefit of using direct metrics estimation compared to entropy measure for TTA and sum measure for SegQC, we compared 3D correction results using these metrics. For 2D slices corrections, we used entropy measure for TTA and sum measure for SegQC, as there are many slices without annotations which we cannot prioritize using 2D metrics estimation based on binary masks.

Fig. 9 shows 3D segmentation correction graphs for fetal body and fetal brain corrections datasets based on either metrics estimation or sum and entropy calculations for SegQC and TTA methods respectively. Comparison was made between: (1) ranking based on SegQC metrics estimation; (2) ranking based on SegQC sum; (3) ranking based on SegQC metrics estimation following estimated error extraction; (4) ranking based on TTA-based metrics estimation; (5) ranking based on TTA-based entropy. Correction performance in 3D for both fetal body and brain segmentations was worse for sum and entropy compared to their metric estimation counterparts of SegQC networks and TTA, respectively

|  |  | Body TTA | | Placenta TTA | |
|---|---|---|---|---|---|
|  |  | \wo | \w | \wo | \w |
| **Dice 3D** | MAE | 0.05 | **0.04** | 0.06 | **0.05** |
|  | Pearson | 0.64 | **0.67** | 0.66 | **0.75** |
| **IoU 3D** | MAE | 0.07 | **0.06** | **0.08** | **0.08** |
|  | Pearson | 0.76 | **0.79** | 0.62 | **0.71** |
| **AVDR 3D** | MAE | **0.06** | **0.06** | **0.20** | 0.24 |
|  | Pearson | 0.56 | **0.57** | 0.38 | **0.53** |
| **Dice 2D** | MAE | **0.06** | **0.06** | **0.09** | 0.10 |
|  | Pearson | 0.70 | **0.71** | 0.61 | **0.64** |
| **IoU 2D** | MAE | **0.08** | **0.08** | **0.11** | 0.12 |
|  | Pearson | 0.75 | **0.77** | 0.65 | **0.67** |
| **AVDR 2D** | MAE | 0.10 | **0.09** | **0.24** | 0.27 |
|  | Pearson | 0.64 | **0.65** | 0.44 | **0.45** |

**Table 5.** Metrics estimation results comparison between SegQC network that was trained without (\wo) and with (\w) augmentations.

**Mask augmentations ablation**. We compared the performance of SegQC networks trained with and without mask augmentations of a single segmentation network on test sets of stratified datasets. Table 5 lists the results. Mask augmentations usually improve metrics estimation results for both fetal body and placenta segmentation quality estimations, with the largest difference observed for the placenta 3D overlap metrics.

## 5. Discussion

Quality control of structures segmentation in volumetric medical images is key for detecting errors in clinical practice and improving model performance in semi-supervised and active learning. This paper presents SegQC, a novel segmentation quality control method based on error segmentation network. Our approach provides quality control in 3D and 2D, as well as detection of error regions, and thus can be used for a variety of important use-cases ranging from scans segmentation quality control to indication of segmentation error regions. Unlike other methods that focus on a single quality estimation metric, our approach computes multiple metrics, including overlap estimates of Dice score and the Intersection over Union (IoU, Jaccard index) metrics, and volume size estimate of the absolute

Relative Volume Difference (ARVD) metric. In addition, we present a novel error extraction method that can be applied on the segmentation error result, which extracts segmentation error regions from the network output for segmentation error regions detection.

The paper presents a new evaluation methodology of segmentation corrections data for quality control estimation. Segmentation corrections data is created by correcting masks of a segmentation network result. Studies show that ground truth segmentation that was created by correcting a segmentation mask is different from the original mask mostly by segmentation error (Chlebus et al., 2019). The original segmentation masks introduce an inductive bias for regions with high observer variability, helping the user to focus on the error regions and making the error regions the most likely to be corrected. In contrast, with the stratified qualities evaluation method, the ground truth quality does not necessarily reflect segmentation errors, but rather a combination of segmentation error and observer variability. Moreover, correction data enables the evaluation of error detection by identifying ground truth error regions and comparing them with the estimated error regions from the QC method.

The drawback of segmentation corrections data evaluation is the limited types of quality that we can assess using this method. We are limited by the quality of the segmentation masks from which the ground truth segmentations were created. Thus, the evaluation on stratified qualities datasets is still important, and evaluation on corrections data can be used as a complimentary evaluation to get a better sense of the pure segmentation error quality estimation.

To better understand the benefits in training a dedicated quality control network, in this work we performed an extensive comparison to the unsupervised TTA based quality estimation method. The quality control evaluation included both traditional metrics such as Pearson correlation and MAE, as well as and additional corrections graph evaluation in 3D and 2D to quantify the ranking capabilities of the quality estimation methods.

Our studies on segmentation corrections data indicate that the supervised SegQC method outperforms TTA-based quality estimation in most cases in terms of the Pearson correlation and the MAE for fetal body and fetal brain structures segmentation. However, the difference in correction performance based on quality estimation ranking is relatively small, indicating that specifically for ranking purposes, using TTA-based evaluation may be enough for the application at hand.

Studies on stratified datasets demonstrate a similar phenomenon for the fetal body segmentation, where SegQC method outperforms TTA for Person correlation and MAE metrics but has only a small improvement for the correction graph evaluation. However, for the placenta segmentation task Pearson correlation and MAE results were mixed, especially for 2D evaluation. This may be due to the high observer variability of the placenta segmentation task, which makes the learning of placenta segmentation quality estimation more challenging.

The results of the segmentation error detection show the benefit of our error extraction method for detecting error regions, even for very small segmentation errors. Our method achieved a recall of 0.77 and a precision of 0.48 for fetal body segmentation error detection task of a high-quality segmentation, compared to a recall of 0.55 and precision of 0.29 without the segmentation error extraction method. For the task of fetal brain segmentation error detection, our method boosted detection performance from a recall of 0.62 to 0.74 and from a precision of 0.52 to precision of 0.55.

The ablation studies with correction graphs evaluation show the benefit of using a direct Dice score estimation compared to entropy and sum estimations for TTA and SegQC methods respectively. They also demonstrate improved Pearson correlation with mask augmentations for the SegQC network.

Our experimental studies have the following limitations. Results indicate a relatively lower quality estimation performance of the SegQC method for 2D compared to 3D data. This may be due to the volumetric nature of the SegQC segmentation network. Fournel et al. (2021) showed the superiority of a 2D regression network compared to 3D network for quality estimation. This can potentially be also the case for the SegQC error segmentation network. Future work can explore the use of a 2D segmentation network for quality estimation. In addition, it can be beneficial to test the performance of difference segmentation network architectures for the error segmentation task.

## 6. Conclusion

This paper describes a new method for segmentation quality estimation of whole scans and for individual scan slices using three different overlap and size metrics. We also presented a method to identify segmentation error regions within scan slices that can help the annotator to focus on error regions. We presented a novel evaluation scheme on segmentation corrections data that distills the segmentation discrepancy caused by error, having a relatively low observer variability. Our method was evaluated on body, brain and placenta structures on corrections and stratified qualities data when applicable.

**Protection of human and animal rights statement**

No animals or humans were involved in this research. All scans were anonymized before delivery to the researchers.

**Conflict of interest**

We declare no conflict of interest.

**Declaration of Competing Interest**

The authors declare that they have no known competing financial interests or personal relationships that could influence the work reported in this paper.

**Data Availability**

The authors do not have permission to share data.

**Acknowledgment**

This research has been partially supported by a grant from the Israel Innovation Authority, Kamin grant 63418 and 72126.